\documentclass[prl,twocolumn,showpacs,groupedaddress]{revtex4}
\usepackage{colordvi,epsfig,color,amsmath,amssymb}

\begin{document}

\title{{\bf Engineered Tunable Decay Rate and Controllable Dissipative Dynamics}}

\author{Zhiguo L\"u $^{1,2}$ \footnote{Electronic address:~\url{zglv@sjtu.edu.cn}}and Hang Zheng $^{1}$}
\affiliation{$^1$Department of Physics, Shanghai Jiao Tong University, Shanghai, 200240, China\\
$^2$Institute of Physics and Department of Electrophysics, National Chiao Tung University, Hsinchu 30010, Taiwan}

\begin{abstract}
We investigate the steering dissipative dynamics of a two-level system (qubit) by means of the modulation of an assisted tunneling degree of
freedom which is described by a quantum-oscillator spin-boson model. Our results reveal that the decoherence rate of the qubit can be significantly suppressed and simultaneously its quality factor is enhanced. Moreover, the modulated dynamical susceptibility exhibits a multi-peak feature which is indicative of the underlying structure and measurable in experiment. Our findings demonstrate that the interplay between the combined degrees of freedom and the qubit is crucial for reducing the dissipation of qubit and expanding the coherent regime of quantum operation much large. The strategy might be used to fight against deterioration of quantum coherence in quantum information processing.

\end{abstract}
\maketitle




%

The emergent field of quantum information processing has spurred research activities on the controlled manipulation of a qubit.
The requisite quantum coherence is very fragile, and lost through interaction with environment or other sources. Such interactions degrade the information and inasmuch as it does exist in nature, how would one implement realistic quantum manipulation at a rate much faster than decoherence occurs? By introducing a dynamical structure with assisted tunneling, we obtain an effective modulation of dissipation, which leads to an engineered decay of the qubit. It avoids the fast loss of coherence of quantum states, which is one of basic prerequisites for quantum information processing. It is applied in different areas of forefront fundamental research and new emerging topics\cite{ladd}.

One of key challenges for solid state qubits is to perform quantum coherent manipulation despite the decoherence resulting from the environment.  For example, generally, a critical element for manipulation is the interdot tunneling barrier of coupled quantum dots (QD), which have been designed and realized in the hetero-structure and other solid state schemes\cite{hayashi,gorman}. However, those are principally set by the geometrical constrictions between the QDs defined in the fabrication\cite{gorman} even if tunnel-barrier transmittances can be modified by gate voltages. Thus, it is an important challenge in experiment to increase the tunneling coupling to preserve coherent dynamics against decoherence due to uncontrollable degrees of freedom\cite{gorman}.

In this paper we propose a powerful efficient mechanism to obtain robust coherent dynamics of a qubit even in a strongly dissipative environment.  A qubit coupled to a harmonic oscillator has been implemented by different systems in experiment, such as an exciton in a
QD coupled strongly to a phonon in an ultrahigh-Q phonon cavity, or a cooper-pair box coupled to a nanomechanical oscillator\cite{brandes,LaHaye,Hohenester}. We propose a realistic setup where assisted-tunneling between qubit states is used to control efficiently the dynamics by coupling elements, a harmonic oscillator acting as a\textit{ controllable} degree of freedom. Specifically, the oscillator can be implemented using the intrinsic lattice vibration of a phonon cavity, i.e., single frequency optical phonons\cite{brandes}, or the collective motion of a designed quantum structure, e.g., a suspended nanomechanical oscillator or carbon nanotube on a gate\cite{huttel,Dai}. Thus, we consider a class of systems with induced tunneling assisted by the oscillator,
\begin{eqnarray}\label{system}
&&H_s=-{\frac{\Delta}{2}}\sigma _{x}+{\frac{\epsilon(t)}{2}} \sigma _{z}+\omega _{0}a^{\dag }a + {\frac{g_{0}}{2}}(a^{\dag}+a)\sigma _{x},
\end{eqnarray}
where $\Delta$ is the tunneling element\cite{note}, $\epsilon(t)$ the bias, $\sigma _{x}$ and $\sigma_{z}$ are Pauli matrices. In QD qubit structures, $\Delta$ is proportional to the overlap of wave functions in the two QDs, $\epsilon(t)$ is the voltage difference between the dots which is tuned to zero in the measurement.  $g_0$ is linear coupling between the qubit and the oscillator, which determines the assisted tunneling strength, while $a$ and $a^{\dag}$ bosonic operators with frequency $\omega_0$. When $\epsilon(t)=0$,  Eq. (\ref{system}) becomes the independent boson model (IBM) with the Hamiltonian $H_{\mathrm{IBM}}=-\Delta\sigma_x/2+\omega_{0}a^{\dag }a +g_{0}(a^{\dag }+a)\sigma _{x}/2$ , describing a two-state system coupled with a dynamic structure. This is a basic model for many physical and chemical processes\cite{mah}.

In order to show how the dynamical fluctuations in the solid-state environment influence the coherent properties of the qubit, the coupling of qubit states to the heat bath is described by the spin-boson model (SBM)\cite{shnirman,rmp,book},
\begin{eqnarray}\label{SBM}
&&H=H_s +\sum_{k}\omega _{k}b_{k}^{\dag }b_{k} +
\sum_{k}{\frac{g_{k}}{2}} (b_{k}^{\dag }+b_{k})\sigma _{z},
\end{eqnarray}
$b_{k}^{\dag }$ ($b_{k}$) is the creation (annihilation) operator of a boson mode with frequency $\omega _{k}$, and $g_{k}$ denotes the coupling constant. In this work an Ohmic bath is considered and its spectral density is characterized by: $\sum_{k}g^2_{k}\delta(\omega-\omega_k)=2\alpha\omega \exp(-\omega/\omega_c)$, where $\alpha$ is the dimensionless coupling constant and $\omega_c$ is the cut-off frequency.

Here employing an analytical approach based on unitary transformations and perturbation theory, we calculate the steered non-Markovian non-equilibrium dynamics, equilibrium dynamics and the susceptibility of the model at $T=0$. The approach puts the qubit's and oscillator's degrees of freedom on equal footing. It works well for the coupling constant $0\leq\alpha<1$ and any bare tunneling $\Delta$, and then can reproduce the well-known results of the SBM [14-21]. In this paper we set $\hbar =1$ and $k_{B}=1$.

Unitary transformations, $H^{\prime\prime}= e^{S_2}e^{S_1} H e^{-S_1}e^{-S_2}$, are applied to $H$ and their aim is to take into account the correlation between the spin and bosons\cite{zhe,sh}. We propose the following form for the generators,
$S_1=\frac{g_0}{2\omega_0}(a^{\dag}-a)\sigma _{x}$ ,
$S_2=\sum_{k}\frac{g_{k}}{2\omega _{k}}\xi _{k}(b_{k}^{\dag
}-b_{k})[\sigma_{-} e^{X}+\sigma_+ e^{-X}] $,
where $\sigma_{\pm}=(\sigma_z\pm i\sigma_y)/2$ and $X=(a^{\dag}-a) g_0/\omega_0$. Here we introduce in $S_2$ a $k$-dependent function $\xi _{k}$ and its form will be determined later. The transformation can be performed to the end and the transformed Hamiltonian are divided into three parts,
$H^{\prime\prime }=H_{0}^{\prime\prime }+H_{1}^{\prime\prime
}+H_{2}^{\prime\prime }$,
\begin{eqnarray}
&&H_{0}^{\prime\prime }=-{\frac{\eta \Delta}{2}} \sigma _{x}
+\omega_0a^{\dag}a+\sum_{k}\omega _{k}b_{k}^{\dag }b_{k}
+\Xi , \\
&&H_{1}^{\prime\prime
}=\sum_{k}V_{k}\left(b_{k}^{\dag}\sigma_{-} e^{X}+b_{k}\sigma_+ e^{-X}\right),
\end{eqnarray}
where the energy shift $\Xi=-\frac{g_{0}^{2}}{4\omega _{0}}-\sum_{k}\frac{g_{k}^{2}}{4\omega
_{k}}\xi _{k}(2-\xi
_{k})$, the renormalized coupling $V_k= g_{k} \frac{\eta\Delta}{\omega _{k}}\xi_k$,
$
\xi _{k}=\frac{\omega _{k}}{\omega _{k}+\eta\Delta }$,
$\eta =\exp \left[-\sum_{k}\frac{g_{k}^{2}}{2\omega _{k}^{2}}\xi
_{k}^{2}\right] $,
and
$H^{\prime\prime}_2=H^{\prime\prime}-H^{\prime\prime}_0-H^{\prime\prime}_1$. $H^{\prime\prime}_0$ is the unperturbed part of $H^{\prime\prime}$ and, obviously, it can be solved exactly in which the spin (qubit), the oscillator and the environment are decoupled. The eigenstate of $H_{0}^{\prime\prime}$ is a direct product: %
$|s\rangle|n_a\rangle
|\{n_{k}\}\rangle$, where $|s\rangle $ is the eigenstate of $\sigma_x$:
$\sigma_x |s_{1}\rangle = |s_{1}\rangle $ and $\sigma_x |s_{2}\rangle = -|s_{2}\rangle $,
$|n_a\rangle$ is the Fock state of the oscillator, and $|\{n_{k}\}\rangle$ is the eigenstate of environment with $n_{k}$ bosons for mode $k$. In particular, $|\{0_{k}\}\rangle$ is the vacuum state in which $n_{k}=0$ for every $k$. The ground state of $H_{0}^{\prime\prime}$ is $|g_{0}\rangle =|s_{1}\rangle|0_a\rangle|\{0_{k}\}\rangle $. $H_{1}^{\prime\prime }$ and $H_{2}^{\prime\prime }$ are treated as perturbation and they should be as small as possible. For this purpose $\xi _{k}$ and $\eta$ are determined to make $H_{1}^{\prime\prime }|g_{0}\rangle =0$ and $\langle g_0| H^{\prime\prime}_2|g_0\rangle=0$, respectively, whose forms are essential in our approach. %

In our method $H^{\prime\prime }_0$ is treated as the unperturbed Hamiltonian, $H^{\prime\prime }_1$ the perturbation which contains the terms of single-bath-boson transition, and $H^{\prime\prime }_2$ is neglected since it contains the terms of multi-bath-boson non-diagonal transitions or the terms of simultaneous transition of bath-bosons and oscillator. In $H^{\prime\prime }_0$, the tunneling has been already renormalized by $\eta$ which comes from the contribution of all diagonal transitions of bath-boson\cite{mah}. %

The equation of motion in the Heisenberg picture for any operator $\sigma(t)=\exp(iH^{\prime\prime }t)\sigma\exp(-iH^{\prime\prime }t)$ is $i\dot{\sigma}=[\sigma,H^{\prime\prime }]$, where the time derivative is abbreviated by a dot. Then, we can derive the chain of equations as follows,
\begin{eqnarray}
\dot{\sigma}_+&=&{\mbox~~}i\eta\Delta\sigma_++i\sum_kV_kb^+_k\sigma_xe^X,\\
\dot{\sigma}_-&=&-i\eta\Delta\sigma_--i\sum_kV_ke^{-X}\sigma_xb_k,\\
i\frac{d}{dt}(b^+_k\sigma_x)&=&-\omega_kb^+_k\sigma_x-V_k\sigma_+e^{-X},\\
i\frac{d}{dt}(\sigma_xb_k)&=&{\mbox~~}\omega_k\sigma_xb_k+V_ke^{X}\sigma_-.
\end{eqnarray}

Quantum fidelity can measure environment-induced decoherence. It is defined as $F(t)=\mbox{Tr}(\rho(0)\rho(t))$.
Thus, suppose the initial state is $|\psi(0)\rangle
=\frac{1}{\sqrt{2}}[|s_1\rangle+|s_2\rangle] |0_a\rangle|\{0_{k}\}\rangle
$ with $\langle\psi(0)|\sigma_z|\psi(0)\rangle=1$ and
$\langle\psi(0)|\sigma_x|\psi(0)\rangle=0$. The average of
the operator $\sigma(t)$ is denoted as $\bar{\sigma}(t)=\langle\psi(0)|\sigma(t)|\psi(0)\rangle$ and we can obtain the integro-differential equations,
\begin{eqnarray}
\dot{\bar{\sigma}}_+(t)-i\eta\Delta\bar{\sigma}_+(t)=-\sum_kV^2_k\int^t_0dt'
\bar{\sigma}_+(t')e^{i\omega_k(t-t')}\mathfrak{F}_{t',t}, \\
\dot{\bar{\sigma}}_-(t)+i\eta\Delta\bar{\sigma}_-(t)=-\sum_kV^2_k\int^t_0dt'
\bar{\sigma}_-(t')e^{-i\omega_k(t-t')}\mathfrak{F}_{t,t'},
\end{eqnarray}
where $\mathfrak{F}_{t',t}=\langle0_a|e^{-X(t')}e^{X(t)}|0_a\rangle$ and $\mathfrak{F}_{t,t'}=\langle0_a|e^{-X(t)}e^{X(t')}|0_a\rangle$ can be evaluated by Feynman Disentangling of operators\cite{mah}. These are non-Markovian. In this work, we pursue controlled non-Markovian transient dynamics. The integro-differential equations can be solved by Laplace transformation and the result for $F(t)=[1+\bar{\sigma}_z(t)]/2=[1+P(t)]/2$ is
\begin{eqnarray}\label{pt}
F(t)
=\frac{1}{2}+\frac{1}{2\pi}\int^{\infty}_{-\infty}\frac{\Gamma(\omega)\cos(\omega
t)d\omega}{[\omega-\eta\Delta-\Sigma(\omega)]^2+\Gamma^2(\omega)},
\end{eqnarray}
where
\begin{eqnarray}
&&\Sigma(\omega)=e^{-\lambda}\sum^{\infty}_{l=0}\frac{\lambda^l}{l!}\sum_k\frac{V^2_k}{\omega-\omega_k-l\omega_0},\\
&&\Gamma(\omega)=\pi
e^{-\lambda}\sum^{\infty}_{l=0}\frac{\lambda^l}{l!}\sum_kV^2_k\delta(\omega-\omega_k-l\omega_0).
\end{eqnarray}
with $\lambda=g_0^2/\omega_0^2$. These equations are the main result of the work. $\Sigma(\omega)$ and $\Gamma(\omega)$ are obtained at $T=0$.
\begin{figure}[t]
\epsfig{file=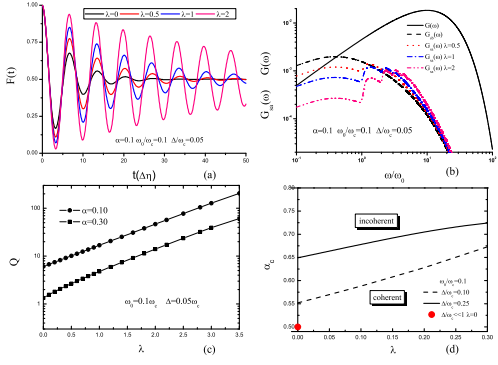,width=85mm}
\caption{\label{fig1}
(a)Time evolution $F(t)$ modulated by the oscillator with different $\lambda$. (b)Spectral density $G(\omega)=\sum_k \frac{g_k^2}{4} \delta(\omega-\omega_k)$ and modulated spectral density $G_{sa}(\omega)=e^{-\lambda}\sum_{l=0}^{\infty}\frac{\lambda^{l}}{l!} \sum_k V_k^2 \delta(\omega-\omega_k-l\omega_0)$. When $\lambda=0$, it become $G_{sb}(\omega)=\sum_k V_k^2 \delta(\omega-\omega_k)$.
(c) Quality factor as a function of $\lambda$ for different dissipative strength. (d)Phase diagram of coherent-incoherent transition in the presence of oscillator tuning. The scaling result $\alpha_c=0.5$ of the SBM ($\Delta/\omega_c \ll 1$) is shown by the red dot.}
\end{figure}

We summarize our results in Fig. \ref{fig1}. One can see a damping oscillation of the SBM at $\lambda=0$ due to dissipation\cite{vol,cos}. Fidelity decays due to interference with the environment. When tuning on the coupling to the oscillator $\lambda> 0$, quantum dynamics exhibits robust coherent behavior even in the strong dissipative regime. Furthermore, as $\lambda$ increases (the oscillator-qubit coupling becomes strong at a fixed $\omega_0$, or the frequency of the oscillator decreases at a fixed $g_0$.),  $F(t)$ shows much weaker damping behaviors and coherence becomes more robust (See Fig. \ref{fig1}(a)).

From Fig. \ref{fig1}(b) it is clear that the coupling to the oscillator has a significant effect on the energy spectrum. $\Gamma(\omega)$ determines the dissipative effects and also relates the spectral density. For the SBM ($\lambda=0$), by using the second-order perturbation theory, we get $\Gamma(\omega)=\pi \sum_k g_k^2 \delta(\omega-\omega_k)/4=\pi G(\omega)$, while from the transformed Hamiltonian (Eqs.(3) and (4)), we obtain $\Gamma(\omega)=\pi \sum_k V_k^2 \delta(\omega-\omega_k)$ and give the effective spectral density $G_{sb}=\Gamma(\omega)/\pi$. By analogy, the modulated spectrum density $G_{sa}$ is defined as $G_{sa}(\omega)=\Gamma(\omega)/\pi=e^{-\lambda}\sum_{l}\frac{\lambda^{l}}{l!} \sum_k V_k^2 \delta(\omega-\omega_k-l\omega_0)$, which has a multi-peak structure with a shoulder in the lower energy in contrast to the bare Ohmic spectrum. As $\lambda$ increases, the shoulder becomes lower and at the same time clear ridges and peaks above the shoulder emerges. It has been known that the low-frequency part of spectrum function determines the dissipative behaviors of qubit\cite{book}. Thus, as the engineered spectrum $G_{sa}$ possesses the less weight of low frequency part it leads to the steered dynamics with dissipative reduction. Therefore, tuning the coupling of the qubit to the oscillator can control the decoherence process of the qubit.

In the Hamiltonian Eq.(\ref{system}), quantum oscillator with an assisted-tunneling channel is introduced to change quantum decoherence. By means of unitary transformations one can see obviously that there happens the mixing ($b_{k}^{\dag}\sigma_{-} e^{X} + h.c.$) between the assisted-tunneling channel of quantum oscillator and the dissipative channel of bosonic reservoir. Thus it results in neutralizing the inevitable decoherence effect caused by the noisy environment. The prefactor $e^{-\lambda}$ in the fidelity dynamics is a signature of this result. Furthermore, the reduction of the decoherence rate originates from the Poisson distribution with the dressing factor $e^{-\lambda}$ in $\Gamma(\omega)$. Therefore, the coherence of the qubit with a dissipative environment can be significantly enhanced and improved.

The quality factor is defined as $Q=\omega_{\mathrm {eff}}/\Gamma(\eta \Delta)$, in which $\omega_{\mbox {eff}}$ is the effective Rabi frequency of the qubit, and $\Gamma(\eta \Delta)$ is the engineered decay rate modulated by the oscillator. Fig. \ref{fig1}(c) displays the quality factor as a function of $\lambda$. As $\lambda$ increases, $Q$ shows a steep increase from 6.09 to 205 for $\alpha=0.1$, a remarkable enhancement by two orders of magnitude. Note that $\omega_{\mathrm{eff}}$ is the solution of the equation $\omega_{\mathrm {eff}}-\eta\Delta-\Sigma(\omega_{\mathrm {eff}})=0$\cite{vol,ks,zhe}. The solution is positive real only when $\alpha<\alpha_c$. For $\alpha >\alpha _{c}$ there is no solution $\omega _{\mathrm {eff}}>0$ so that $\alpha =\alpha _{c}$ determines the critical point corresponding to a coherent-incoherent transition. For $\lambda=0$, it is easy to obtain the well-known result $\alpha_c=1/2$ in the scaling limit $\Delta \ll \omega_c$ \cite{zhe}. Figure \ref{fig1}(d) shows the phase diagram of the coherent-incoherent transition. It is clear that the coherent regime becomes much broader with increasing $\lambda$. In other words, the oscillator provides both a degree of freedom to steer the qubit dynamics and also a broad parameter space to efficiently manipulate its coherence.

Another important quantity is the susceptibility\cite{egg,ks,and},
\begin{eqnarray}\label{chi}
&&\chi ^{\prime \prime }(\omega )=\frac{1}{2\pi}\int_{-\infty }^{\infty }dt e^{i\omega t}
\langle [\sigma _{z}(t)\sigma _{z}-\sigma _{z}\sigma
_{z}(t)]\rangle_H ,
\end{eqnarray}
where $\langle ...\rangle_H=\mbox{Tr}[\exp(-\beta
H)...]/\mbox{Tr}[\exp(-\beta H)]$. The correlation function is
calculated at $T=0$ as follows.
\begin{eqnarray}
&&\langle\exp(iHt)\sigma _{z}\exp(-iHt)\sigma _{z}\rangle_H \approx
\left(\langle e^{i\tilde{H}t}\sigma_{-}e^{-i\tilde{H}t}\sigma_{+}\rangle_{\tilde{H}} \right.  \nonumber\\
&&\left.+\langle e^{i\tilde{H}t}\sigma _{+}e^{-i\tilde{H}t}\sigma_{-}\rangle_{\tilde{H}} \right) \mathfrak{F}_{t,0}.
\end{eqnarray}
Here we assume that in the transformed Hamiltonian $H_a=\omega_0a^{\dag}a$ is uncorrelated with $\tilde{H}=H''-H_a$. The correlation functions can be calculated by means of Green's function and the result is
\begin{eqnarray}
&&\langle e^{i\tilde{H}t}\sigma_{-}e^{-i\tilde{H}t}\sigma
_{+}\rangle_{\tilde{H}}\langle
e^{-X(t)}e^{X}\rangle_{H_a}   \\
&&={1\over\pi}\int^{\infty}_0d\omega\frac{\gamma
(\omega)}{[\omega-\eta\Delta-R(\omega)]^{2}+\gamma
^{2}(\omega)}e^{-i\omega t} e^{-\phi(t)}, \nonumber
\end{eqnarray}
where $\phi(t)=\lambda(1-e^{-i\omega_0 t})$. Then, it is substituted into Eq.(\ref{chi}) and for $\omega\ge 0$,
\begin{eqnarray}\label{chi2}
\chi''(\omega)&=&\frac{e^{-\lambda}}{\pi}\sum^{\infty}_{l=0}\frac{\lambda^l}{l!}  \\ \nonumber
&&\times \frac{\gamma (\omega-l\omega_0)}{[\omega-l\omega_0
-\eta\Delta-R(\omega-l\omega_0)]^{2}+\gamma ^{2}(\omega-l\omega_0
)} ,
\end{eqnarray}
where $R(\omega)=\sum_k\frac{V^2_k}{\omega-\omega_k}$ and $\gamma(\omega)=\pi\sum_kV^2_k\delta(\omega-\omega_k)$.
For $\lambda=0$, $\chi''(\omega)=\pi^{-1} \gamma
(\omega)\{[\omega-\eta\Delta-R(\omega)]^{2}+\gamma
^{2}(\omega)\}^{-1}$ is the result of SBM\cite{bulla,zhe}, while for
$\alpha=0$, $\chi''(\omega)=e^{-\lambda}\sum_{l} \lambda^l
\delta(\omega-l\omega_0-\Delta)/l!$ is that of IBM\cite{mah}.

\begin{figure}[t]
\epsfig{file=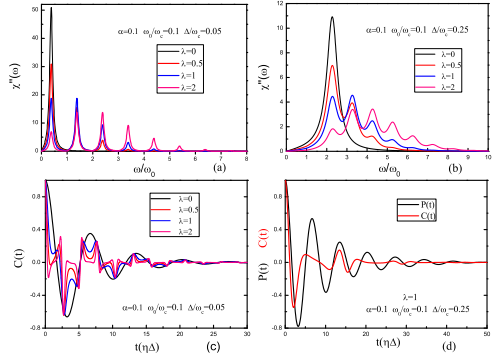,width=85mm}
\caption{\label{fig2} (a) and (b) $\chi''(\omega)$ as a function of $\omega/\omega_0$ for different $\lambda$. (c) The correlation function $C(t)$.  (d) $C(t)$ vs $\bar{\sigma}_z(t)\equiv P(t)$. Note C(0)=1=P(0) is numerically exact.}
\end{figure}

Figure \ref{fig2} shows clearly the Poisson distributed character of the susceptibility spectrum [Figs. \ref{fig2}(a) and \ref{fig2}(b)]. Without the coupling to the oscillator, the spectrum has a Lorentzian shape with a single peak near $\eta\Delta$ which is a well known aspect of the SBM\cite{bulla,cos}. Turning on the coupling to the oscillator, one can see that a multi-peak structure emerges, in which there is a finite frequency interval $\omega_0$ between any two nearest-neighbor peaks, which is the intrinsic frequency of the oscillator. Moreover, the weight of every peak would be redistributed with increasing $\lambda$. Specifically, the position of the highest peak will shift, at the same time the height and width of every peak also change. All these features in the spectrum can be detected and confirmed experimentally in the optical spectrum of QDs. In Fig. \ref{fig2}(c), we plot the symmetrized equilibrium correlation function $C(t)=\langle\{\sigma_z(t),\sigma_z\}\rangle/2$, where $\{A,B\}=AB+BA$ \cite{and}. As $\lambda$ increases, the oscillatory character shows up chorally multi-peak structure in the susceptibility because $C(t)$ and $\chi''$ satisfy the correlation $C(t)=\int_0^{\infty}d\omega\chi''(\omega)\cos(\omega t)$. In the absence of dissipation ($\alpha=0$), our result immediately recovers the exact result of the IBM\cite{mah}. Besides, without the coupling to oscillator($\lambda=0$), the dynamics $C(t)$ illustrates a dissipative behavior with a decay rate $\gamma(\eta\Delta)$ \cite{zhe}, and $C(t)=\bar{\sigma}_z(t)$ in this case\cite{egg}. While the dissipative and controllable components coexist, $C(t)$ has a much different character from $P(t)$, namely $C(t)\neq \bar{\sigma}_z(t)$ [Fig. \ref{fig2}(d)]. One of important reasons is the different initial preparation conditions for two physical quantities\cite{book}. In particular, $C(t)$ is an equilibrium correlation function while $\bar{\sigma}_z(t)$ needs an initial preparation process and then performs a non-equilibrium time evolution. From a mathematical viewpoint, the kernel of $\bar{\sigma}_z(t)$ accounts the total effect of summing multi-boson ($l \omega_0, l=0,1,2...$) self-energy in Eq.\ref{pt}(quantum coherent superposition), resulting in damped oscillations with the corresponding modulated decay rate. The kernel of $C(t)$, on the other hand, considers the contribution of each multi-boson term independently, and $C(t)$ takes account of the summation of separate term as a Poisson distributed weight in Eq.\ref{chi2}(incoherent superposition). Therefore, $C(t)$ is different from $\bar{\sigma}_z(t)$.

In summary, we study the controllable dissipative dynamics coherently tuned with a harmonic oscillator using an analytical approach based on unitary transformations. We show that the modulated decay rate is controlled and its quality factor improved under suitable steer of the oscillator. In other words, when the oscillator is tuned appropriately, the coherence of the system exhibits becomes significantly more robust. The scheme is realistically done, from an experimental point of view, i.e. the oscillator can be modeled by a nanostructural oscillator, a quantum beam, an optical phonon mode, or even a cavity mode. The mechanisms discussed in this paper may be used in designing and constructing qubit-manipulation tools to preserve coherence, and also in finding applications to the dynamics of light-harvesting complexes and quantum information transfer\cite{milburn,pc}. The results of this work hold for zero temperature. The issue of decoherence for finite temperature is interesting and in progress.

Z.G. L\"u is grateful to M.P. Blencowe, J.J. Lin, C-H. Chung, C. Chu and R. Winkler for discussions. The work was supported by the NNSF of China (Grants No.10734020 and No.10904091), NBRPC (Grant No.2011CB922202).

{\rm \baselineskip 20pt}


\begin{thebibliography}{99}
\bibitem{note} {The transition is due to spontaneous tunneling, whereas the physical process of the fourth term in $H_s$ is known as structure-assisted(stimulated) tunneling.}
\end{thebibliography}
\end{document}